# Altmetrics of "altmetrics" using Google Scholar, Twitter, Mendeley, Facebook, Google-plus, CiteULike, Blogs and Wiki

Saeed-Ul Hassan, Uzair Ahmed Gillani

saeed-ul-hassan@itu.edu.pk

Information Technology University, 346-B Ferozepur Road, Lahore (Pakistan)

**Abstract:** We measure the impact of "altmetrics" field by deploying altmetrics indicators using the data from Google Scholar, Twitter, Mendeley, Facebook, Google-plus, CiteULike, Blogs and Wiki during 2010- 2014. To capture the social impact of scientific publications, we propose an index called alt-index, analogues to h-index. Across the deployed indices, our results have shown high correlation among the indicators that capture social impact. While we observe medium Pearson's correlation ($\rho$= .247) among the alt-index and h-index, a relatively high correlation is observed between social citations and scholarly citations ($\rho$= .646). Interestingly, we find high turnover of social citations in the field compared with the traditional scholarly citations, i.e. social citations are 42.2% more than traditional citations. The social mediums such as Twitter and Mendeley appear to be the most effective channels of social impact followed by Facebook and Google-plus. Overall, altmetrics appears to be working well in the field of "altmetrics".

**Introduction**

In scholarly world, altmetrics are getting popularity as to support and/or alternative to traditional citation-based evaluation metrics such as impact factor, h-index etc. (Priem et. al., 2010). The concept of altmetrics was initially proposed in 2010 as a generalization of article level metrics and has its roots in the #altmetrics hashtag (McIntyre et al, 2011). In recent years, social media has dramatically impacted research workflow - scholars prefer to discuss and share their work on social media platforms like twitter with hashtags to reach relevant audience, Facebook and Google-plus to share it with their social circle or beyond. In addition, scholars have started extensively using mediums like Mendeley to organize social references and CiteULike to bookmark papers related to their interest (Haustein & Siebenlist, 2011; Nielsen, 2007).

Naturally scholars want to remain updated in their field of research and take interest in latest articles only, but existing citation process need time to accrue, thus, this is what motivates scholars to move towards altmetrics domain (Priem & Berge, 2010). Altmetrics is basically count of citations or mention over social media platforms (i.e. Twitter, Facebook, Mendeley etc.) and considered as an alternative to existing scholarly measurements. Since a new publication may appear more rapidly on the social media platforms, e.g. it is more reasonable to expect tweets and blogs mentions related to any article on its publication day or within one month of its publication. Hence social media platforms are considered to be a valuable tool for scholars trying to promote their research publication in less time. We have number of websites providing altmetrics tracking with free and paid plans (e.g. impactstory.com, altmetrics.com, etc.)

Alternative metrics or commonly called as "altmetrics" are usually build on information from social media usage, and we could employ it side-by-side with citation-based metrics. Since the inception of "altmetrics", a vast group of scholars are debating that altmetrics could deliver information about the impact on diverse communities like clinicians, practitioners, and the general public as well as help to track the use of diverse research products like datasets, software and blog posts (Brody et al, 2006). Given the amount of efforts being invested in this field, it becomes extremely important to measure the impact of "altmetrics". This paper provides a brief overview of "altmetrics" field by deploying known altmetrics indicators. In addition, we also propose a measure of alt-index, analogous to traditional h-index, which captures the social impact of scholarly publications.

The rest of the article is organized as follows. In the next section, we discuss related studies. We then proceed with explaining the altmetrics indicators we use for this study. Further we present a case study of scholars in the field of "altmetrics". Finally, we present conclusions along with directions for future work.

**Literature Review**

Usage of citation metrics has significant importance for the measurement of research impact, but this method has some inherited limitations. The process of finding citation latency of even high impact articles is a bit time taking process and usually first citations appear in 1-2 years or even longer (Pavitt, 1985). Citations are normally used to measure only one type of research products while neglecting other important emerging datasets such as: Google+, Facebook and Twitter. The shortcoming of citation remains a major issue from the ages. The research administrators are trying from decades to create more realistically diverse measure of research impact, painstakingly gathering diverse indicators including patents (Pavitt, 1985), acknowledgements (Cronin & Overfelt, 1994), doctoral committee membership and many other (Sugimoto et al, 2008).

Traditional ways to find scholarly impact was to check citation of a document. As mentioned earlier citation calculation was a lengthy process and normally took longer time to get cited. Methods were explored to find scholarly impact of social media content by using different altmetrics. They have gathered information from different social media sources relevant to authors, research topics and research dimensions and determined the scholarly impact of documents using social media analytics (Priem et al, 2012).

In earlier days, it was a time taking process to find relationships in bibliometric research parameters. Relationship among different bibliometric research parameters was found using Social Network Analysis. The main goal had been to determine social structure of bibliometric and its impact, so relationship between different actors and bind them in one network was established to fulfill this need (Al et. al, 2012). Social media content is evolving as one of the major resource in evaluation of impact in multiple industries. User generated content is a useful medium to find its impact and evaluate any particular sector. Social contents may also help governments to govern their policies based on user views by analyzing positive and negative feedback expressed in social content. Moreover it may also help governments to check their popularity with these analytics and evaluate where they can perform somehow better (Grubmuller et al, 2013).

Social media analysis and techniques of textual analysis also help to evaluate quality of user generated content regarding any particular topic. For instance, finding quality and impact of user generated content through social media by analyzing multiple resources played a vital role in past. For a baseline work is initiated with yahoo answers and evaluated quality of content using its main attributes (i.e. question likes, thumbs up, thumb down, no of comments) and evaluated quality of content discussed in that particular topic (Agichtein et al, 2008) . Twitter is playing a major role in social media analytics as it is providing facilities of micro blogging with huge impact. Twitter corpus is used for analysis and opinion mining of data by creating their own classifier to process data. Majorly they use data cleaning, data processing and analysis to find useful patterns for making meaningful predictions (Pak et al, 2010) .

Massive use of social content creates room for researchers to find interesting patterns. Text messages generated by social media was a major breakthrough as it revealed another way to analyze scholarly impact by analyzing relevant discussion related to research topic. Yang et al (2014) created a neural network model created named as self-organizing map that creates clusters on the basis of sentiment and keywords. Using these clusters association between message contents and sentiments of message based on keywords was discovered.

Social media has created its ripples in field of education too. Brody (2006) proposed an algorithm to evaluate sentiment of teachers feedbacks for students based on defining sentiment value for keywords. Whenever teacher provides some feedback related to any student this algorithm executes that feedback content against its defined weightage and evaluate a general sentiment about the feedback. Finally the feedback points from all teachers are accumulated and a general sentiment is proposed based on sentiment score. Among numerous proposed factors, researchers also worked to propose t-factor in comparison to h-index, they gathered around 69 tweets and its retweets form Scopus database and calculated t-factor of a researcher (Bornmann & Haunschild, 2015).

**Data & Methodology**

**Approach**

The goal of the research design was to provide a brief overview of "altmetrics" field by deploying known altmetrics indicators. We also devised a measure of alt-index, which captures the social impact of scientific publications. Using the keyword "altmetrics" on Google scholar[1], we collected data of 77 scholars working in the field of "altmetrics". Among the selected group, we found that only 47 scholars were active on social media platforms, thus, we restricted our dataset to these 47 scholars only. Further, we collected social citations and scholarly citations data of the publications produced by the selected scholars during 2010 - 2014.

As we were more concerned to find the social presence of scientific publications, so we started by identifying the core indicators from the literature that can be helpful to answer our research goals. Initially we collected data about tweet mentions, twitter follower, twitter retweets, Facebook likes, Facebook shares, Facebook mentions, Google plus mentions, Google 1+, LinkedIn mentions, linked in shares, Mendeley reader, CiteULike bookmarks, wiki mentions, blogs mentions, reddit mentions. But when we started collecting data related to scientific publications, we identified some indicators that do not provide any data correspond to the publications in our dataset. So we started with the indicators that responded to at least one publication. The final data set was 217 publications that responded to 16310 events against the selected social indicators. Table 1 presents the final indicators along with their data sources.

---

[1] https://scholar.google.com/

**Table 1:** Indicators for data collected for this study.

| Data Source | Indicator(s) |
|---|---|
| Twitter | Mention Count, Favorites, Retweets |
| Facebook | Mention Count, Likes, Shares |
| Google Plus | One-plus |
| Mendeley | Reader |
| CiteULike | Bookmarks |
| Blogs | Mentions Count |
| Wiki | Mentions Count |

*Twitter mentions, favorites and retweets:* All counts of tweets mention, favorites and retweets were collected from impactstory, altmetrics, and twitter manually.

*Facebook mentions, likes and shares:* All counts of Facebook mentions, likes and shares were collected from impactstory, altmetrics and Facebook manually.

*Google Plus mentions and one plus:* All counts of Google plus mentions and one plus were collected manually from impactstory, altmetrics and Google plus.

*Mendeley reader count:* Mendeley readers were counted using impactstory, altmetrics and Mendeley.

*CiteULike count:* CiteULike bookmarks were counted using impactstory and altmetrics.

*Blog & Wiki count:* Blog and Wiki counts were counted using impactstory and altmetrics.

As each paper was linked with multiple indicators, it was required to propose some formula to get some accumulative value of social citation that we can use for comparing with its scholarly citation. We assign score to each indicator based on its importance, i.e. all those indicators which were directly citing any publication were given 1 score, while those who were linking to just liking or reading were give half importance as compared to cited ones. A detail scoring of each indicator is shown in Table 2. Once we get value of all social citations, final dataset contains 47 researchers, 214 publications, 5238 scholarly citations and 7452 social citations as shown in Figure 1.

**Table 2:** Weightage of indicators.

| INDICATOR | SCORE |
|---|---|
| Twitter Mention | 1 |
| Facebook Mention | 1 |
| Google Plus Mention | 1 |
| Blog Mention | 1 |
| Wiki Mention | 1 |
| Mendeley Readers | 0.5 |
| CiteUlike Bookmark | 1 |
| Twitter Favourites | 0.5 |
| Facebook Likes | 0.5 |
| One Plus | 0.5 |

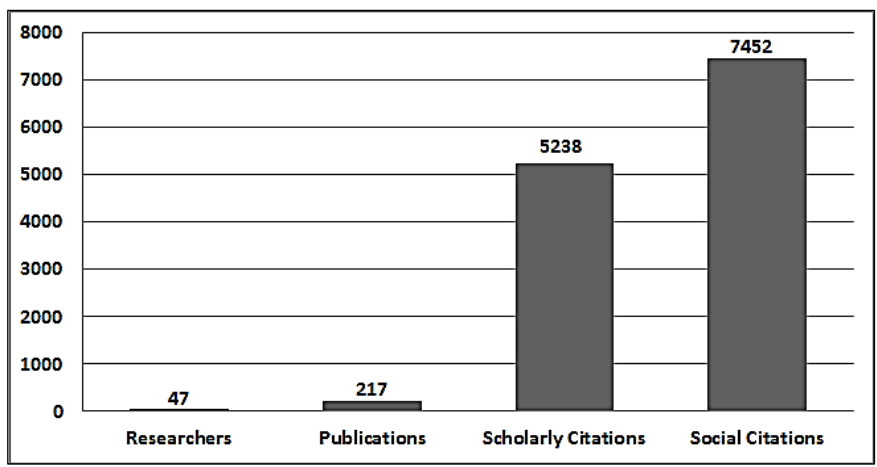

**Figure 1:** Researchers, publications, scholarly citations and social citations during 2010-2014.

After having all the basic dataset, now we wanted to address our basic research question to propose an index on the basis of social citation value, which should be a valid altmetric for our research evaluation and help us to bridge some scholarly gap. We named it alt-index and calculated it using same formula we have for h-index, the only difference between h-index and alt-index is that it is based on social citation values rather than scholarly citation. So formula for alt-index goes as:

> *"if x number of publications have at least x social citations, its*
>
> *alt-index will be x. "*

We calculated social index of all researchers based on this newly proposed alt-index, we will discuss all these results in detail in results and discussion section.

**Data and Code Availability**

Data and scripts are available on Github: https://github.com/slab-itu/altmetrics-dataset.

**Results and Discussions**

In this section we present that how our approach will address our research questions:

1. Can we use social citations as an alternative to traditional citations?
2. Can we propose an index based on social presence of a researcher which can compete with existing h-index?
3. Evaluate impact of researchers working in field of "altmetrics" using altmetrics.

At first we examine yearly trend of publications, scholarly citations and social citations using a line graph during 2010 – 2014, as shown in Figure 2. As expected publications in earlier years of our selected time window receive more citations since they have a larger time window, however, in contrast to scholarly citations, we find that papers published in recent years receive high social citations.

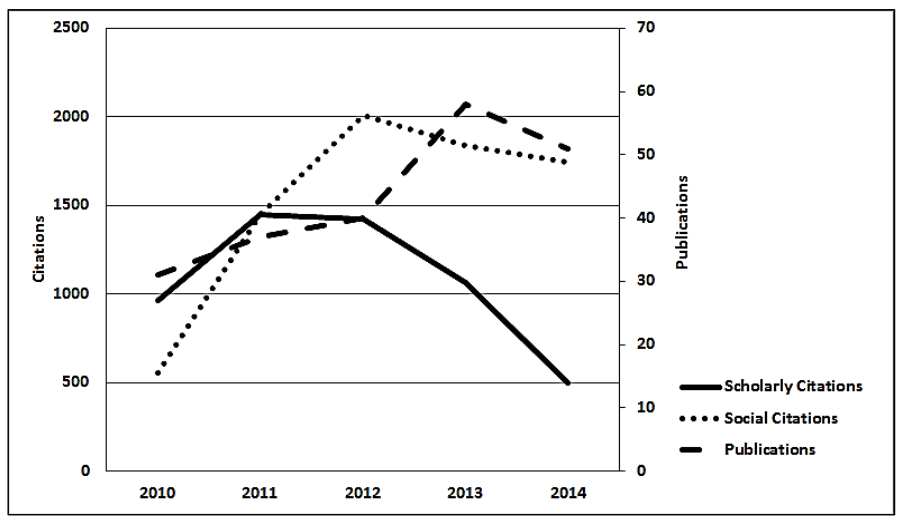

**Figure 2:** Publications, scholarly citations and social citations over years.

Further, in order to visually investigate the relationship between scholarly citations and social citations received by the publications in our dataset, we mapped these both indices in Figure 3, it shows positive correlation ($\rho = .646$), as around 10% articles showed equal weightage of social and scholarly citation value. While 40% have difference of around 4 points, the rest of the articles showed difference of at max 10 points between scholarly and social citation value.

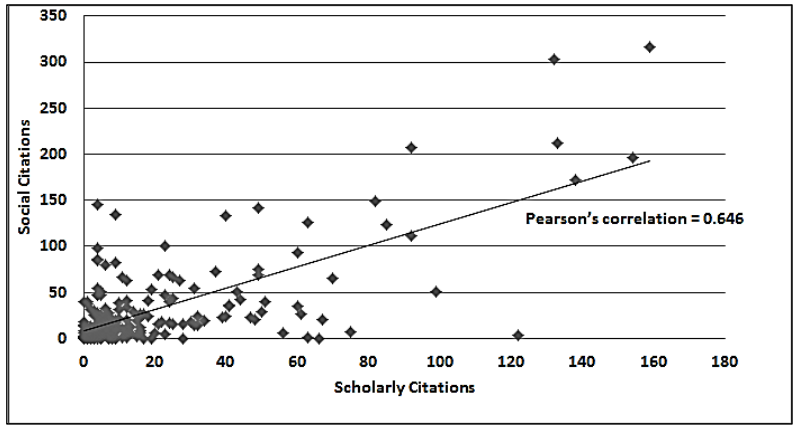

**Figure 3:** Social citations vs. scholar citations of selected researchers.

In Figure 3, we find some high peaks of social citation value, further to find that which indicators contribute high towards social citations; we map percentage of all these indicators on a horizontal bar graph as shown in Figure 4. Interestingly we find that major contribution of social citations comes from Mendeley and Twitter, followed by Facebook.

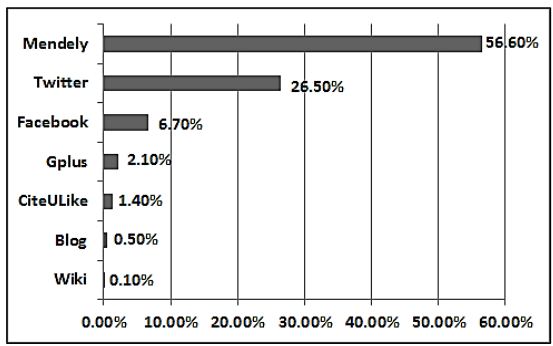

**Figure 4:** Social indicators drill down for all social citations.

Further to investigate the relationship between social citations and scholarly citations we compute the Pearson's correlation coefficient of scholarly citations, social citations, h-index, alt-index and publications of the researchers in our dataset during 2010-2014 (see Table 3). We find positive correlation between h-index and alt-index (ρ= .247). As discussed above, we find relatively high correlation between social citations and scholarly citations (ρ= .646). Interestingly, publications show relatively high correlation (ρ= .613) with social citations as compared to scholarly citations (ρ= .425). Similarly, we find relatively high correlation (ρ= .601) between publications and alt-index as compared to publications and h-index (ρ= .431). This may reflect the behavior of researchers in the field that they have high tendency to discuss their research outcomes on social media platforms. Note that Table 4 shows Pearson's correlation coefficient of all social indicators at paper level.

**Table 3:** Pearson's correlation coefficient of indices

|  | scholarly citations | social citation | h-index | alt-index | publications |
|---|---|---|---|---|---|
| scholarly citations | 1.000 | .646 | .676 | .215 | .425 |
| social citation | .646 | 1.000 | .529 | .570 | .613 |
| h-index | .676 | .529 | 1.000 | .247 | .431 |
| alt-index | .215 | .570 | .247 | 1.000 | .601 |
| publications | .425 | .613 | .431 | .601 | 1.000 |
| a. Listwise N = 45 | | | | | |
| **. Correlation is significant at the 0.01 level (2-tailed). | | | | | |

Finally we plot h-index and alt-index of selected researchers overall during 2010-2014 (See Figure 5). While the researcher R27 (Mike Thelwall) shows highest h-index value in the selected duration, researcher R14 (Heather Piwowar, co-founder of impactstory) show highest alt-index value in the field. Drilling down further, Figure 6 shows yearly trends of publications, scholarly citations and social citations of 12 researchers that receive highest social citations. We see that across the community, researchers receive high social citations to the papers published in recent years compared with the scholarly citations to the recent papers.

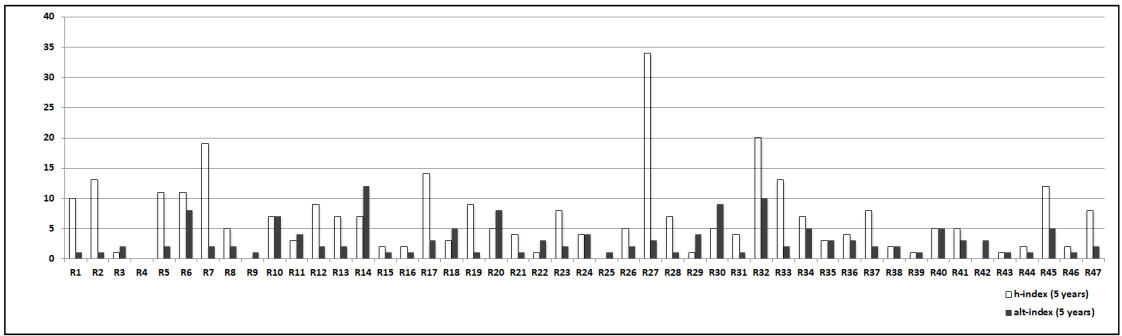

**Figure 5:** H-index and alt-index of selected researchers during 2010 to 2014

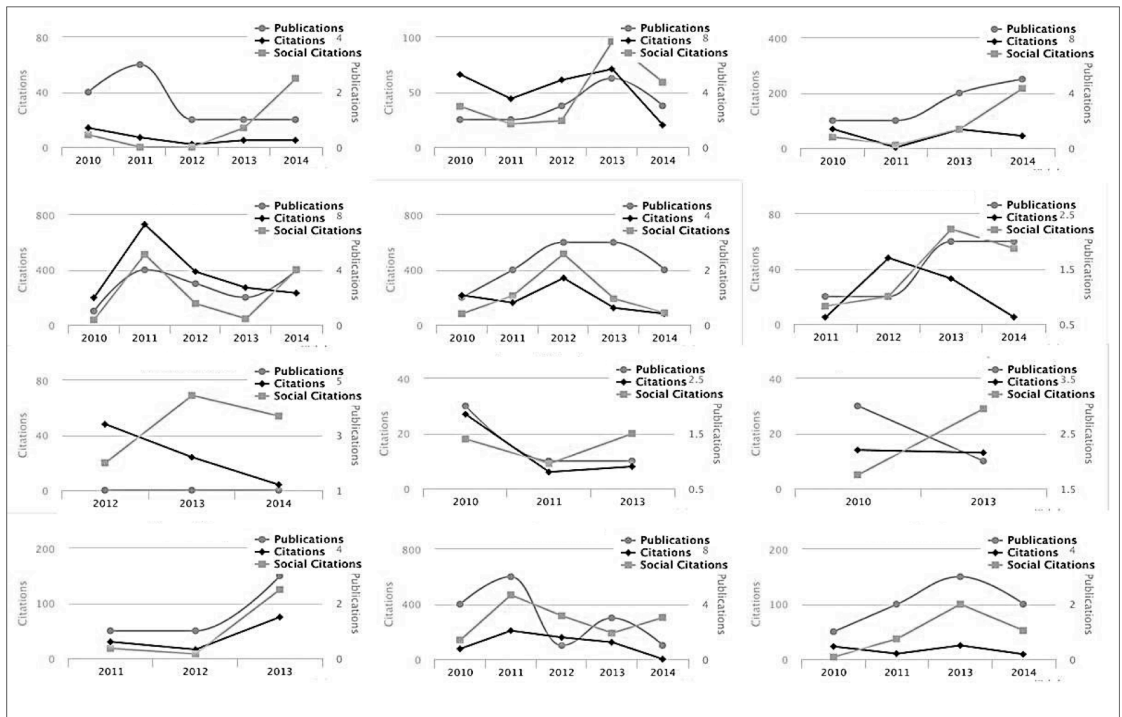

**Figure 6:** Publications, citations and social citations of selected researchers during 2010 to 2014

**Concluding remarks**

In this paper we measure the impact of "altmetrics" by deploying altmetrics indicators. In addition, we also propose a measure of alt-index which captures the social impact of scientific publication. Our results have shown high correlation among the indicators that capture social impact. Similarly, we have observed medium correlation among the alt-index and h-index. Interestingly, we find high turnover of social citations in the field compared with the traditional citations during 2010- 2014, i.e. social citations are 42.2% more than traditional citations. The social medias such as Twitter and Mendeley appears to be the most effective channels of social impact followed by Facebook and Google-plus. Overall, altmetrics appears to be working well in the field of "altmetrics". In future, we plan to include semantic analysis of social media contents along with the scientific publications produced by the scholars. We also plan to build an online tool to show the scholarly activity of researchers on social media channels. Note that the data gathered in this research is publically available - we encourage other community members to build more sophisticated models using this dataset.

Table 4: Pearson's correlation coefficient of social indicators

| | tweets | Facebook | Google | Blog | Wiki | Shares | Like | Retweets | One_plus | Favorites | Bookmarks | Readers | Scholarly citations | Social citations |
|---|---|---|---|---|---|---|---|---|---|---|---|---|---|---|
| tweets | 1.000 | .452** | .548** | .343** | .272** | .336** | .369** | .727** | .373** | .654** | .421** | .312** | .179** | .689** |
| Facebook | .452** | 1.000 | .353** | .324** | .087 | .406** | .681** | .285** | .282** | .341** | .221** | .234** | .149* | .421** |
| Google | .548** | .353** | 1.000 | .196** | .307** | .307** | .282** | .446** | .697** | .450** | .231** | .246** | .117 | .496** |
| Blog | .343** | .324** | .196** | 1.000 | .182** | .111 | .161* | .155* | .137* | .092 | .369** | .334** | .196** | .361** |
| Wiki | .272** | .087 | .307** | .182** | 1.000 | .303** | .058 | .194** | .123 | .168* | .369** | .208** | .137* | .257** |
| Shares | .336** | .406** | .307** | .111 | .303** | 1.000 | .443** | .400** | .304** | .340** | .236** | .191** | .157* | .341** |
| Like | .369** | .681** | .282** | .161* | .058 | .443** | 1.000 | .360** | .223** | .423** | .177** | .214** | .193** | .380** |
| Retweets | .727** | .285** | .446** | .155* | .194** | .400** | .360** | 1.000 | .266** | .810** | .274** | .263** | .156* | .604** |
| One_plus | .373** | .282** | .697** | .137* | .123 | .304** | .223** | .266** | 1.000 | .295** | .210** | .138* | .115 | .336** |
| Favorites | .654** | .341** | .450** | .092 | .168* | .340** | .423** | .810** | .295** | 1.000 | .256** | .264** | .171* | .604** |
| Bookmarks | .421** | .221** | .231** | .369** | .369** | .236** | .177** | .274** | .210** | .256** | 1.000 | .423** | .301** | .464** |
| Readers | .312** | .234** | .246** | .334** | .208** | .191** | .214** | .263** | .138* | .264** | .423** | 1.000 | .555** | .830** |
| Scholarly citations | .179** | .149* | .117 | .196** | .137* | .157* | .193** | .156* | .115 | .171* | .301** | .555** | 1.000 | .502** |
| Social citations | .689** | .421** | .496** | .361** | .257** | .341** | .380** | .604** | .336** | .604** | .464** | .830** | .502** | 1.000 |

**. Correlation is significant at the 0.01 level (2-tailed).
*. Correlation is significant at the 0.05 level (2-tailed).
c. Listwise N = 217